\documentclass{article}
\usepackage[utf8]{inputenc}
\usepackage[T1]{fontenc}

\usepackage{blindtext}
\usepackage{subfig}
\usepackage[toc,page]{appendix}
\usepackage{booktabs}
\usepackage{float}

\title{Towards Evaluation of Autonomously Generated Musical Compositions: \\
A Comprehensive Survey}
\author{\textbf{Daniel Kvak}\\
	Faculty of Arts\\
	Masaryk University\\
	Brno, Czech Republic\\\\
	ORCID: 0000-0001-7808-7773}

\usepackage{natbib}
\usepackage{graphicx}

\begin{document}

\maketitle
\begin{abstract}
    There are many applications that aim to create a complete model for an autonomously generated composition; systems are able to generate muzak songs, assist singers in transcribing songs or can imitate long-dead authors. Subjective understanding of creativity or aesthetics differs not only within preferences (popular authors or genres), but also differs on the basis of experienced experience or socio-cultural environment. So, what do we want to achieve with such an adaptation? What is the benefit of the resulting work for the author, who can no longer evaluate this composition? And in what ways should we evaluate such a composition at all? 
\end{abstract}


\newpage
\section{Introduction}
The scope of algorithmic composition includes a wide range of tasks, which consist of the creation of melodies, chords, rhythms or even lyrics, i.e., all the typical components of music. The purpose of this work is to present the output of the neural network without any modifications. The aim of the research is not to romanticize artificial intelligence and its use in algorithmic composition, but on the contrary to point out its advantages and disadvantages. We therefore start from the idea that only undistorted results carry a certain aesthetic, which is destroyed by later authorial adaptations. There are many applications that aim to create a complete model for an autonomously generated composition; systems are able to generate muzak songs, assist singers in transcribing songs or can imitate long-dead authors. But why? This is the right question that we would like to address in this paper. Subjective understanding of creativity or aesthetics differs not only within preferences (popular authors or genres), but also differs on the basis of experienced experience or socio-cultural environment. So, what do we want to achieve with such an adaptation? What is the benefit of the resulting work for the author, who can no longer evaluate this composition? And in what ways should we evaluate such a composition at all? Let's try to paraphrase musicologist Matěj Kratochvíl here: \textit{“So far I have not found anyone who would publicly consider whether the world will somehow enrich the existence of a new composition by a dead author, which in comparison with his actual work sounds as if it was composed by a machine. So, I dare say that the music world has not gained anything new.”} \citep{kratochvil_2020}

\section{Analytical evaluation of the created work}
One possibility is analytical evaluation, which would examine the diversity of tones or counterpoints using statistical methods, and then compare these findings with our learning corpus. Even this method is not able to capture a certain aura of the work; how the generated composition will be evaluated by artists, musicologists and ordinary listeners.

\subsection{Evaluation using a tone attractor}
We can find certain satisfaction in Eduard R. Miranda \citep{10.1162/106365604773955120}, who in his text describes the only factor that an artificially generated composition must meet in order to be described as creative: surprise. He illustrates this idea with a model based on the virtual world of artificial composers and critics: while composers have the task of creating melodies, critics evaluate these melodies and then decide with whom to link reproduction. The algorithm uses a virtual form of selective pressure to support the evolution of suitable composers of melodies. This model draws inspiration from the bird world, where some singers use melodies to attract a mating partner. Although this model uses heuristics of evolutionary algorithms, we can apply similar procedures to current programs based on deep neural networks. As already mentioned in the introduction, the problem of autonomous musical composition falls among the time series models; this applies not only to models based on artificial intelligence, but also to genetic algorithms.\\
\\
The interesting thing about this model is that in order to achieve a high evaluation score, the composer must first raise expectations, then violate this assumption. Thus, there is a constant tension between the expected sequence and the moment of surprise, but only surprising transitions are highly valued. This selective process ensures that the genotype representing the highest quality "tone attractor" advances to the next generation. The taxonomy of the offspring is selected at random and one third of the population is subsequently removed. This model clearly uses Darwin's principle of \textit{survival of the fittest}, which can support the development of coherent melody repertoires in the virtual world of agents. The problem with this model may be the strictness of critics who only blindly follow a set of probabilities and their deviations using the coefficients of the Markov chain.

\section{Aesthetic and musicological evaluation of the created work}
The reception of music is influenced by a number of factors, and these factors are also subject to an objective but also subjective nature. The perception of the components of the composition is often derived from personal experiences and feelings, which only further complicates the possibility of aesthetic evaluation. \citep{schindler2017measuring} Since it seems almost impossible to find a method that would fully take into account the general subjective evaluation of the artistic values of musical works, we are offered a questionnaire survey, which could verify based on lay and professional interest, whether the composition follows certain rules of musical composition, or whether it is even possible to label the resulting composition as a creative act. In the case of automated systems for algorithmic composition, which are to reproduce human creativity in some way, there is a problem with the evaluation of their performance: the concept of artistic creativity escapes a formal, unambiguous and effective definition. This makes it difficult to evaluate these systems quite thoroughly.

\subsection{The problem of defining computational creativity}
So how do we define creativity in this case? I am aware that defining and clarifying the concept of creativity would be an extremely difficult, but rather unrealistic task. Certainly, there are many textbooks that generally describe creativity as a set of skills that enable artistic, scientific, or other creative activity. \citep{petrova1999tvovrivost} Other sources define creativity as a phenomenon during which something new and valuable arises, whether in tangible (invention, literary, painting) or immaterial (idea, scientific theory, musical composition) form. \citep{mumford2003have} Psychologist Jana Petrová distinguishes between inventive and innovative creativity: \textit{“Innovative creativity manifests itself in proposing new unusual and at the same time socially desirable ideas, thoughts, solving problems that occur in the organization and in discovering problems, shortcomings or opportunities. Carriers of invective creativity are constructors, designers, rationalizers, improvers and inventors. Innovative creativity is manifested in the implementation of a new idea. Its essence is in understanding the new idea, in identifying with it and in its further creative elaboration and elaboration with regard to the problems in its transformation into material reality.”} \citep{petrova1999tvovrivost} However, the aim of this work is not to define creativity. with a few more definitions that can help us validate the generated songs.

\subsection{Evaluation using the theory of musical metacreaction}
Although the topic of creativity would certainly deserve its own master's thesis, we will focus only on the issue of musical metacreaction, which falls under the young field of computational creativity. Generating a musical composition is one of the most difficult tasks for artificial intelligence. Approaches based on machine or deep learning have already been able to convincingly mimic output on shorter sections of music, but most current models cannot create longer coherent compositions. \citep{porter_2019} Musical metacreaction focuses on the analysis and evaluation of creative tasks, which include composition, interpretation, improvisation, accompaniment, mixing and more. This field covers all dimensions of the theory and practice of computational generative music systems, from artistic approaches to purely scientific ones, including discourses on topics falling under the socio-cultural sciences. \citep{pasquier2017introduction} Computational creativity differs from traditional, rationally defined problems related to artificial intelligence in that it often solves problems for which the notion of solution optimality is difficult to define. These problems include those for which there are no clearly defined states of victory, defeat or goal, no obvious and complete objective functions, or no defined preferential relations. \citep{pasquier2017introduction}
\\
\\
Following the open problem from the time series, an alternative description and distribution of creativity according to Bown et al. to generative and adaptive creativity. \citep{bown2015manifesto} Generative creativity goes beyond human activities and includes any process in which new objects can arise, both material and intangible. Adaptive creativity is defined as a deliberate cognitive act: adaptive is in the sense that it improves the situation of the actor. This difference is crucial for the ideas of computational creativity, as there are many human processes at different socio-cultural levels where the creative product may not contain any obvious value. According to this theory, creative results can therefore be considered as a combination of generative and adaptive creative processes. \citep{bown2015manifesto} Our emerging model, which aims to imitate Antonín Dvořák's music, could therefore be considered creative to some extent. However, we can look at systems of algorithmic composition as objects of scientific and artistic research, which is why it seems important to approach them in appropriate ways. What evaluation methods, therefore, Pasquier et al. in the article \textit{An Introduction to Musical Metacreation} suggests? \citep{pasquier2017introduction}\\
\\
•	\textbf{Authors}. Artists, designers and computer scientists, as the authors of the model, are usually the first to evaluate the generated composition.\\
•	\textbf{Users, peers and experts}. Composers, musicians or sound designers can provide important critical feedback.\\
•	\textbf{Audience}. Many creative models are designed to create works that will be presented to a wider audience. However, the question remains whether measuring the popularity of the output has any value for the aesthetic and musicological validation of the work.\\
•	\textbf{Press and media coverage}. The interest of critics or reviewers may represent some evaluation, but as in the previous case, it is difficult to talk about the value of such validation.\\
•	\textbf{Peer reviewers, curators and jurors}. The output can also be evaluated from the point of view of the academic review procedure, which guarantees considerable neutrality and professional assessment.\\
•	\textbf{Theoretical and analytical measures}. The emergence of critical studies that respond to the work can offer a broader view of the professional public.\\
•	\textbf{Empirical studies}. Qualitative or quantitative research, which will appeal to laymen and professional musicians, can be a compromise solution for model validation.\\
\\
Although the list includes both academic and popular validation options, these procedures do not offer a specific and universal way in which we could evaluate compositions. The very understanding of the concept of creativity is often subject to subjective perceptions, opinions or experiences, so it seems inevitable to leave the question of perception of creativity to the respondents themselves in an empirical study.

\subsection{Evaluation using a questionnaire survey}
Although the questionnaire survey is not the most suitable method for evaluating musical compositions, it offers us the opportunity to obtain more specific data from a wider group of evaluators. However, the heterogeneity of such a group is considerably limited; Although the questionnaire was shared between lay listeners and professional musicians, this group still falls into a certain social bubble that feels a certain inclination towards musical composition. The homogeneity of the group is disturbed at least by the fact that some respondents are familiar with this project, while others are not.\\
\\
The most suitable solution seems to be a semi-structured questionnaire, which can be approached as qualitative and quantitative research. Open-ended questions offer space to communicate personal preferences, or at least to understand the broader context in the respondent's answers, while closed-ended questions are designed to provide us with some analytical certainty that is loosely tied to open-ended questions. In the questionnaire survey, we intentionally ignore the inclination and experience with musical composition, as our goal is aesthetic evaluation, not the evaluation of procedures and changes in these resulting works.\\
\\
•	Please listen to these short examples. What is your opinion on these songs? (open question)\\
•	Do you think that Antonín Dvořák is the author of these compositions? (standardized question)\\
•	Would you distinguish a composition composed of artificial intelligence from a human creation? (standardized question)\\
•	What do you think is creativity? (open question)\\
•	Do you think that artificial intelligence can be creative? (standardized question)\\
•	Now that I mention that these songs were created by artificial intelligence, please try to describe your opinion again. (open question)\\
\\
Respondents were contacted through a simple internet form between January 22 and 29, 2021, and the answers to the standardized questions were definitely yes / rather yes / rather no / definitely no / I don't know. A total of 25 unique respondents completed the questionnaire. All the addressed respondents are members of the interest group \textit{Composers CZ / SK}, so they are united mainly by their interest in musical composition. But what conclusions can we draw from this survey? In order to answer this question, we must first consider the extent to which we can evaluate the aesthetic aspect of the resulting compositions. Individual answer sheets can be found in the diploma thesis appendix.\\
\\
So, let's go back to the introduction and remember the methods of musicological evaluation according to Fiebrink and Caramiaux: control, structure, creativity and interactivity. \citep{fiebrink2016machine} In the next few paragraphs, we will focus on the individual parts of this evaluation, perhaps only with the exception of interactivity, which we cannot apply due to the form of the model. The answers to the first question differ in many ways, so we might assume that we observe a strong influence of the subjective grasp of the musical composition. Several respondents mention the austerity and simplicity of the compositions, which is typical for algorithmic creation, both on the basis of stochastic modeling and using neural networks. While in some opinions the idea prevails that the compositions, despite maintaining a solid or static structure, seem to be imaginative, while others mention the unsystematic or even disharmony of the output. These views do not change at all even after the respondents found out that it was a musical composition created by artificial intelligence. A repeatedly mentioned phenomenon, which does not contribute to the credibility of the output, is the lack of dynamics of the compositions. We will return to this issue in the next chapter. However, some respondents also find a certain form with the compositions of Antonín Dvořák, but rather in the form of slight hints, which take the form of simple motifs. Out of the total number of 25 respondents, 10 stated that Antonín Dvořák is certainly not the author of the mentioned compositions, another 10 leaned towards the answer rather no, while only 2 people thought that the famous composer could create these songs.
\begin{figure}[h]
\includegraphics[width=1\textwidth]{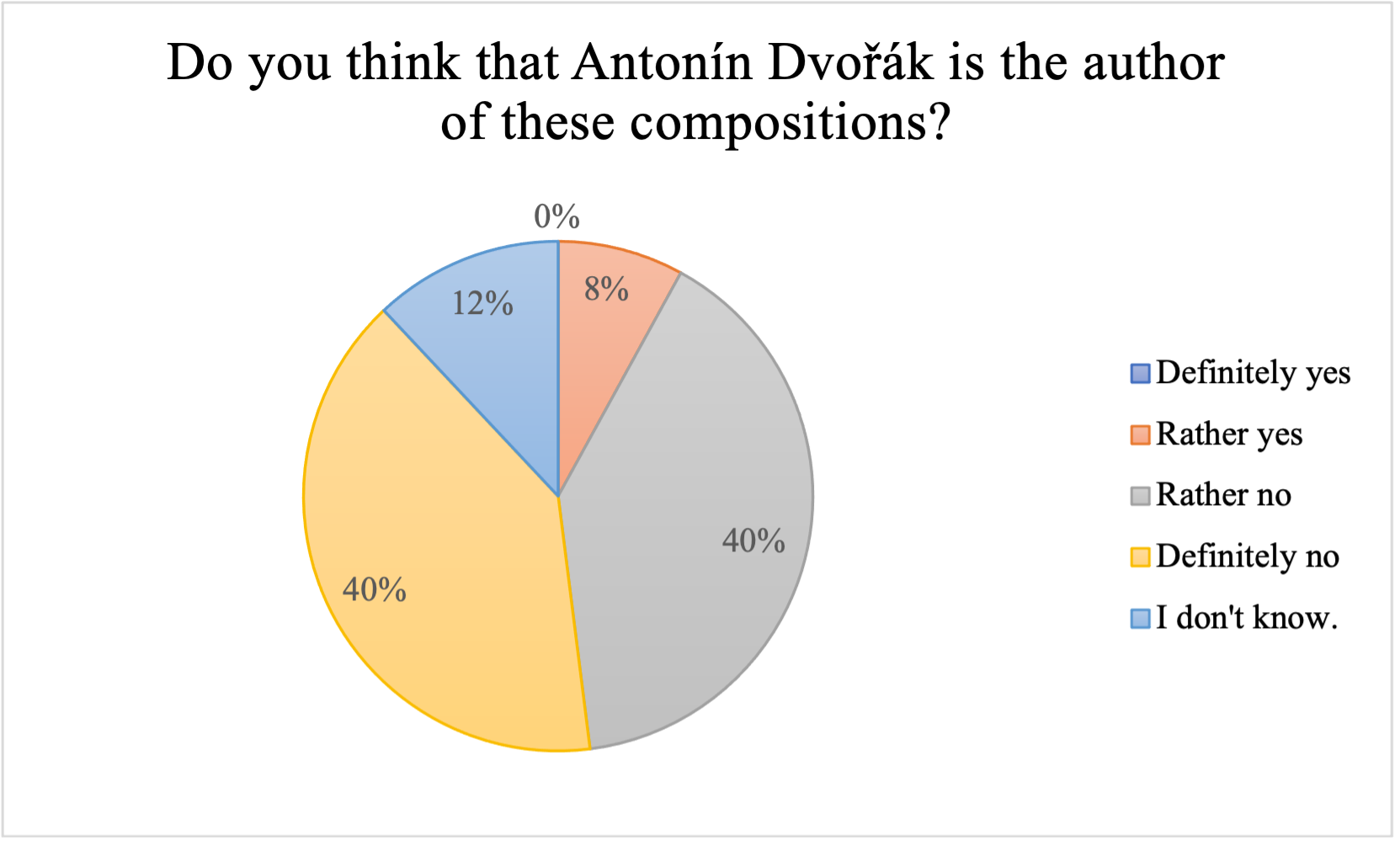}
\centering
\caption{\textbf\scriptsize{A pie chart representing the answers to the question "Do you think Antonín Dvořák is the author of these compositions?"}}
\end{figure}
\\
\\
We will now focus only on the part of the semi-structured survey that deals with the issue of creativity. Even for our respondents, it was not easy to deal with the general definition of creativity. No wonder, however, despite the best efforts of philosophers, psychologists, cognitive scientists and educators, unfortunately there is no generally accepted definition of creativity. Quite interestingly, but seemingly generally, creativity was defined by developmental psychologist Howard Gardner: \textit{“No person, act or product is creative or uncreative in itself.”} \citep{gardner1995creativity} By no means do we want to consider this wording as a dogma, but it allows us to focus on the evaluation of the resulting compositions, not just on examining the various definitions. So, let's leave the definition of creativity to the respondents themselves, and at the same time let us ask whether they consider artificial intelligence to be a truly creative entity.\\
\\
An interesting probe into the subjective nature of music composition evaluation is the connection between the questions „What do you think creativity is?“ And „Do you think that artificial intelligence can be creative?“ Thanks to this phenomenon the decision to apply to some extent to their own definitions of creativity. Let us mention here another, no less popular and just as controversial, division of creativity. Cognitive science professor Margaret Boden divides creativity into three components: \\
\\
•	\textbf{Exploratory creativity} includes a given conceptual space in which creative activity takes place. Exploratory creativity is rather the result of a persistent search in space (e.g.: writing poetry, painting, musical composition).\\
•	\textbf{Transformational creativity} is based on the deliberate transformation of conceptual space, by rejecting or replacing some of its components. Transformational creativity is also responsible for technological innovations (e.g.: internet, mobile phone, car).\\
•	\textbf{Combinatorial creativity} is closely related to transformational creativity. The main phenomenon associated with this type is the connection of elements into new relationships and combinations. \citep{boden2004creative}\\
\\
Among the most common concepts we can find in the table of answers is creativity as a process of creation, either through one's own ingenuity or a combination of already existing elements. A certain imaginary intersection between the respondents and the Boden division can be found in the definition of creativity as a process of transformation or combination \citep{boden2004creative}, but some answers also mention the exploratory phenomenon of creativity (finding context beyond the rules or the ability to assess the current environment).
\begin{figure}[h]
\includegraphics[width=1\textwidth]{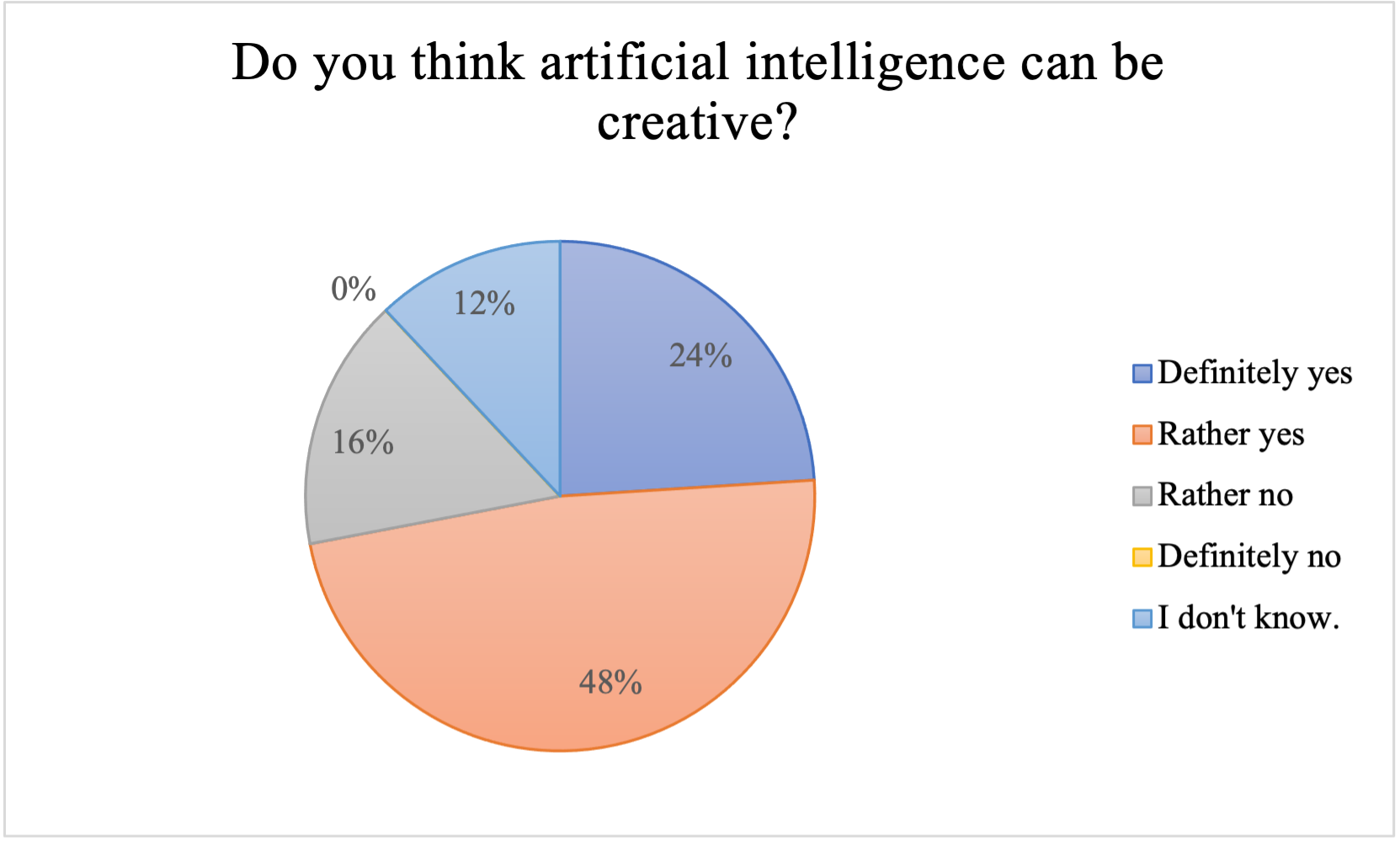}
\centering
\caption{\textbf\scriptsize{A pie chart representing the answers to the question "Do you think artificial intelligence can be creative?"}}
\end{figure}
\\
\\
An interesting phenomenon is the high proportion of respondents who believe that artificial intelligence can exist as a creative entity. It would be extremely difficult to determine what methods of bias each individual answer hides (e.g.: a lay understanding of artificial intelligence, experience with similar models), so we have to look at the data with a significant distance. Positive answers may indicate a high percentage of acceptance of computational creativity, but they may also indicate that respondents do not have a realistic idea of the possibilities of artificial intelligence. This critical view emphasizes what we did not learn, rather than providing us with a probe into what we found. However, we must keep in mind that it would be courageous to verify the question of computational creativity only on the basis of vague definitions, yet the questionnaire survey provides an interesting piece into a complex mosaic of aesthetic validation and offers us a unique insight into the broader context.\
\subsection{Evaluation using the Turing test}
In 1950, the British mathematician Alan Turing wrote an article called \textit{Computing Machinery and Intelligence} \citep{turing2009computing}, in which he presented the first variant of the Turing test, which is still widely discussed. The original model, known as \textit{The Imitation Game}, contains a human observer who communicates with two agents via a text interface. The first agent is a human, while the second agent is a machine. If the observer cannot successfully distinguish man from machine through discourse, then, according to Turing's view, machine has achieved intelligence. \citep{harnad2003minds}\\
\\
It must be added, however, that Turing does not define thinking or intelligence in his work, nor does he claim that passing this test proves idea or intelligence. In addition, the machine agent can deceive the interviewer: mathematical questions can be answered with programmed errors, answers can be intentionally delayed simulating human computation times. According to Steven Harnad, a cognitive scientist, Turing's goal can be considered an epistemic rather than an onic point, while the result is heuristic rather than demonstrative. \citep{harnad2003minds} Alan Turing left important test parameters open, including the number of investigators and agents, their qualifications and the length or organization of the tests. However, the Turing test can be repeated and then averaged to arrive at the induction requirement. Other views suggest that a simplified version without a comparison between two agents, involving only a computer agent and an observer, is satisfactory. \citep{harnad2003minds} \\
\\
Although many different methods have been proposed for evaluating computational creativity, none of them can be easily and generally applied to machines and humans; at least not in a way that would not provoke controversy. The solution to measuring computational creativity against human standards seems obvious: we ask respondents to listen to machine-generated compositions, and we can declare an algorithmic composition system as a creative entity if evaluators cannot separate these compositions from compositions composed by real composers. \citep{belgum1988turing} Without being aware of this fact, the respondents became part of this simple variant of the Turing test. As the primary goal of this paper is the evaluation of autonomously generated compositions, in the interest of generalization, the original authorial compositions of Antonín Dvořák were converted into a symbolic representation of the ABC format. This decision has the effect of reducing the dynamics, pace and other limitations that were set out in the introduction to this work. Of course, there is also the opposite possibility, in which we would leave artificially composed and original compositions at the mercy of realization by real artists, but this experiment would contradict the starting point where we refuse to alter results that carry austere, computational aesthetics.
\begin{table}[]
\centering
\begin{tabular}{@{}llll@{}}
\toprule
Correct answer & Respondent 1 & Respondent 2 & Respondent 3 \\ \midrule
B              & \textbf{B}   & A            & A            \\
A              & \textbf{A}   & B            & B            \\
A              & B            & B            & B            \\
B              & \textbf{B}   & A            & \textbf{B}   \\
B              & A            & A            & A            \\
Success rate:  & 60 \%        & 0 \%         & 20 \%        \\ \bottomrule
\end{tabular}
\caption{\label{tab:table-name}Evaluation using the musical Turing test.}
\end{table}
\\
\\
The reduction of musical compositions to pure, unaltered data output is an interesting step that points to a phenomenon that Professor Joanna Zylinska calls \textit{crowdsourced beauty}: people can appreciate art if it is similar to something they already know. \citep{zylinska2020ai} Forms of generative art using pattern recognition methods are mere mimicry that masks original, truly creative works. \citep{zylinska2020ai} However, if we return to the evaluation methods using Fiebrink and Caramiaux \citep{fiebrink2016machine}, we reveal an interesting phenomenon: if we reduce the dynamic changes and improvisational activities of the performer to the minimum necessary, our respondents can hardly distinguish between original and autonomously generated composition.\\
\\
The social notion of musical creativity thus includes not only the arrangement of tones but above all the wide use of human factors. However, this phenomenon contradicts the validation of the works of art themselves. Although we often witness confident statements from the societies involved, we cannot deny that the primary research goal has been reduced to the mere question of whether a work can pass. To some extent, we accept the fact that we can consider human creativity in part as a computational act. But can this idea be contradicted? Art is one of the few domains for which there was a belief that it will forever remain the prerogative of man \citep{harnad2003minds}, the very mathematical nature of the musical composition, but also of other creative circles, suggests the exact opposite. Curator Jérôme Neutres is of the opinion that generative examples of artificial intelligence are a kind of human-mechanical system: \textit{“Naturally, artists do not construct these machines just to get 'help' but rather to probe the limits of the human idea of creativity and of human-machinic assemblies. These works are thus described as collaborations between the artists and the robotic systems those artists have designed.”} \citep{zylinska2020ai} The method, which can to some extent be called the musical Turing test, seems to be valid only when the generative composition system tries to imitate previously created works. \citep{ariza2009interrogator} But what would we do if artificial intelligence created a unique, truly innovative work of art that is not just an empty box or imitation of existing originals? Is it possible to judge such autonomously generated music according to human standards at all?\

\section{Conclusion}
Although we have proposed several different methods of analytical, musicological and aesthetic evaluation, Matěj Kratochvíl's rhetorical question from the introductory part of this paper \citep{kratochvil_2020} was far from answered. Although the evaluation methods of musical metacreaction offer several possibilities by which we can to some extent determine the impact of similar generative projects (see media evaluation and audience evaluation), the measurability of such outputs is at least problematic. So, let's try to think again about Matěj Kratochvíl's statement and verify the contribution that generative art can provide.
The emergence of models that focus on generative forms of art has brought with it, above all, the loss of illusions that art will remain the exclusive domain of man. However, generative art, and therefore artificial intelligence in general, faces many difficulties. In their reports, journalists often describe the resulting works in layman's terms without any knowledge of cultural or scientific understanding, which further damages the authenticity of these works. The idea of artificial intelligence as omniscient robots, which could one day exterminate and replace humanity, is fueled not only by media companies, often by commercial popularity, which aims to convince investors why their model is the real, unadulterated artificial intelligence. For laymen, AI is a difficult topic to grasp, which is most often associated with the film hero Terminator, or with a similar theme of cult sci-fi films.
We encounter the same situation in the case of computational creativity. The media explosion of artificial intelligence hardly distinguishes between machine learning and the supposed, nowadays non-existent general artificial intelligence. Firms engaged in the application of neural networks are then to some extent forced to maintain this unrealistic picture of AI, which today represents a cultural fetish rather than an object of scientific interest. If, in the case of supposed, truly omniscient artificial intelligence, we speak of the loss of illusions, then this issue points to a perhaps even more interesting perspective of creative tendencies.\\

\bibliographystyle{agsm}
\bibliography{references}
\begin{figure}[p]
\includegraphics[width=1\textwidth]{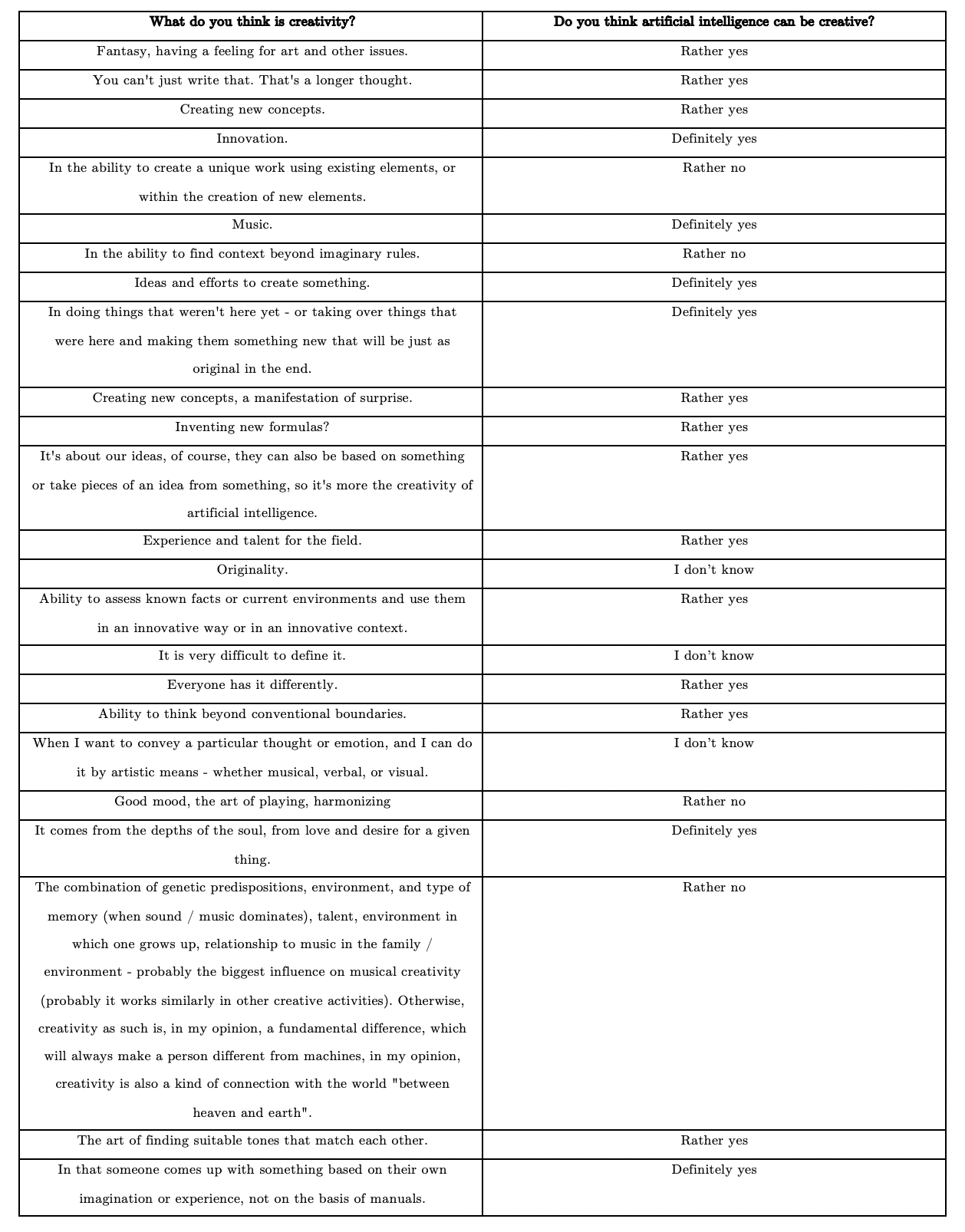}
\centering
\caption{\textbf\scriptsize{Social definition of creativity according to respondents.}}
\end{figure}
\end{document}